\documentclass[twocolumn,showpacs,preprintnumbers,amsmath,amssymb,10pt,aps,prb]{revtex4-1}
\usepackage{epsfig,amsopn}
\usepackage{graphicx}
\usepackage{epstopdf}
\usepackage{amsmath,amssymb}
\usepackage{amsthm}
\usepackage{enumerate}
\usepackage{xcolor}

\begin{document}
\title{Non-local control of spin-spin correlation in finite geometry helical edge}

\author{Sonu Verma}
\affiliation{Department of Physics, Indian Institute of Technology Kanpur, Kanpur 208016, India}
\author{Arijit Kundu}
\affiliation{Department of Physics, Indian Institute of Technology Kanpur, Kanpur 208016, India}
	
\begin{abstract}
An infinite edge of a quantum Hall system prohibits indirect exchange coupling between two spins whereas a quantum spin-Hall edge prohibits out-of-plane coupling. In this study we analyze an unexpected breakdown of this behaviors in a finite system, where the two spins can interact also via a longer path that traverses the whole perimeter of the system. We explain this using an analytical model as well as using tight binding models in real space. Based on this finding, we propose how using a lead far away from the spins can switch the coupling on and off among them non-locally.
\end{abstract}
	
\maketitle
	
\emph{Introduction.}--- Non-local control of the interaction among spins has been a field of intense study in past few years in the end to assist quantum information processing~\cite{RKKYQC} as well as in spintronic applications. Effective interaction among localized spins mediated by the underlying delocalized electrons is described by the Ruderman-Kittel-Kasuya-Yoshida (RKKY) theory~\cite{RKKY}. Controlling such coupling non-locally, such as by optical means~\cite{RKKYTerahz,RKKYOptical},  external magnetic field~\cite{RKKYMagnetic} or applied electric field~\cite{RKKYElectrical1,RKKYElectrical2} among others~\cite{RKKYQdots1,RKKYQdots2,RKKYQdots3,RKKYQdots4,RKKYQdots5,RKKYQdots6,RKKYSC,RKKYMechanical} has been proposed and also some of them has been verified experimentally~\cite{RKKYExpt1,RKKYExpt2,RKKYExpt3,RKKYExpt4}. Among solid-state based systems, spin-orbit coupled systems~\cite{RKKYSO1,RKKYSO2,RKKYSO3}, especially, quantum spin-Hall systems are among the significant candidates that can mediate long-range controllable coupling~\cite{RKKYQSH1,RKKYQSH2,RKKYElectrical1} among spins.

Quantum Hall (QH) and Quantum spin-Hall (QSH) states are characterized by the non-zero spin-Chern number and have topologically protected \textit{chiral} edge states, where a given spin mode can traverse in a given direction~\cite{QSH}. QH states break time reversal symmetry and have edge states where both spins move in chiral channels in the same direction. Whereas, in QSH, the edge states have oppositely moving channels for opposite spins, preserving the time reversal symmetry of the system. Due to the chiral nature of states and the one dimensionality, it is expected that such edge states would carry long range correlation also among two spins placed on the edge, which is indeed what has been explored in recent studies~\cite{RKKYQSH1,RKKYQSH2,RKKYSi}.

The spin-momentum locking (helicity) of the edge states give rise to vanishing out-of-plane RKKY coupling for the spins on a QSH edge, whereas in QH edge, all components of the RKKY coupling vanishes~\cite{RKKYQSH1,RKKYQSH2}. In this work we analyze and propose to manipulate an unexpected breakdown of this result when the geometry is finite. This behavior is a result of the fact that in a finite geometry the helical edge states can come back by traversing the whole edge of the sample. Further, such long coupling between the spins is found to be anti-ferromagnetic in nature and the amplitude of the coupling becomes almost independent of the positions of the spins. We show this using lattice simulation with two models, one in hexagonal lattice~\cite{Haldane,KaneMele} another in a square lattice~\cite{BHZ}. This surprising behavior can also be explained using an analytical model of the edge states. The longer path of interaction between the spins through the whole perimeter of the system can be controlled by using a lead, attached to the edge far away from the spins, which can induce de-coherence in the edge states resulting in turning off the relevant interaction among the spins. This mechanism allows to have a truly non-local control of the coupling between the spins, where none of the local parameters are modified.

\emph{RKKY by infinite chiral edge.}--- The hallmark of the topologically non-trivial states are the chiral edge states where a given spin can move in a definite direction. These states are also protected from back-scattering (without flipping their spins) through a bulk band gap. In particular, the helical edge (running along the $\pm x$ direction) of the QSH phase can be represented by the Hamiltonian $H_{0} = -iv_F \sigma_z \partial_x$, where $\sigma$ is the Pauli matrix of the real spin and $v_F$ is the Fermi velocity. The corresponding Green's function is block-diagonal in up and down spin sector~\cite{RKKYQSH1}:
\begin{align}\label{eq:Gupdn}
G^{\text{QSH}}_{\pm} (x,x';\omega) = -\frac{i}{v_F} e^{i\frac{\omega}{v_F}(x-x')}\theta(\pm(x-x')),
\end{align}
where the $\pm$ refers to up/down spin states respectively. This particular form of the Green's function is a result of spin-momentum locking in the helical edges.

Spin-susceptibility of the system can be captured by looking at the effective spin-spin correlation of two impurity spins mediated by the states of the system. Considering the impurity spins ($\mathbf{S}_1,\mathbf{S}_2$ at positions $x_1,x_2$ on the edge) couple to the delocalized electrons in the edge through the Kondo coupling $H' = -\frac{J}{2}\sum_{r,\sigma,\sigma'} \psi^{\dagger}_{r\sigma}(x)(S_{1,\sigma\sigma'}\delta(x-x_1)+  S_{2,\sigma\sigma'}\delta(x-x_2))\psi_{r\sigma'}(x)$, where $r$ could be left or right moving states, second order perturbation in $H'$ gives the effective RKKY interaction among the impurity spins,
\begin{align}
H_{\text{RKKY}} &= -\frac{J^2}{\pi} \int_{-\infty}^{E_F} d\omega \text{Tr}[(\mathbf{S}_1.\mathbf{\sigma})G(\mathbf{r}_{12};\omega + i0+)(\mathbf{S}_2.\mathbf{\sigma})\nonumber\\
&~~~~~~~~~~~~~~~~~~~~~~~~~~\times G(-\mathbf{r}_{12};\omega + i0+)]\label{eq:rkky}\\
& \equiv \sum_{i,j = x,y,z} \mathcal{J}_{ij} S_{1i}S_{2j}\label{eq:corr},
\end{align}
where $\mathbf{r}_{12}$ is the separation of the two spins and resultant $\mathcal{J}_{ij}$ forms the spin-spin correlation matrix. Eq.~(\ref{eq:Gupdn}) immediately results in vanishing \textit{Ising} interaction among spins which are aligned as up and down spins, i.e, $\mathcal{J}_{zz} = 0$. The appearance of the theta functions with opposite sign in Eq.~(\ref{eq:Gupdn}) is essentially responsible for this behavior, which dictates that $|G_{\sigma\sigma}(x,x';\omega)|$ is non-vanishing only for $x-x'>$ and $<0$ for up and down spin modes respectively. For a QH phase, similar argument follows and as both spins can move in only one direction, the argument of the theta function in Eq.~(\ref{eq:Gupdn}) have the same sign, which results in all correlations $\mathcal{J}_{ij}$ vanishing in a QH edge.

\emph{Lattice simulation.}--- To study a finite topological system, below we consider a Hamiltonian in hexagonal lattice that can represent QH, QSH or normal insulator for different ranges of parameters. A square lattice system has also been studied and detailed in the Appendix \ref{app:I}. The Hamiltonian on the hexagonal lattice reads as~\cite{KaneMele,Haldane}:
\begin{align}
H =& -t \sum_{\langle i,j\rangle,\sigma}  c_{i\sigma}^\dagger c_{j\sigma} + \sum_{\langle\langle i,j\rangle\rangle \sigma}i\lambda_{\sigma} \nu_{ij} c_{i\sigma}^\dagger c_{j\sigma}+ \sum_{i\sigma}\Delta_ic_{i\sigma}^\dagger c_{i\sigma}~, \label{eq:H0}
\end{align}
where $c^{\dagger}_{i\sigma}$ is the electronic creation operator at site $i$ with spin $\sigma$. $t$ is the nearest neighbor hopping, that also serves as our unit of energy. Next to nearest neighbor hopping amplitude, $\lambda_{\sigma}$, is the spin-orbit coupling strength, $\sigma$ represents spins with $\sigma=\pm$ for up/down spin electrons respectively. $\Delta_i =\mu \pm \Delta$ contains the chemical potential $\mu$, which we keep at zero and the staggered potential $\Delta$, where $\pm$ applies to A and B sub-lattices. $\nu_{ij}$ is ${\pm1}$ depending on clockwise or anti-clockwise hopping. This Hamiltonain can be realized in various solid-state systems like silicene, germanene and stanene~\cite{Ezawa2011a,Konschuh2010}. When $\lambda_{\sigma} = \sigma\lambda$, the Hamiltonian is time-reversal symmetric (and break inversion symmetry) and the ground state is a QSH state when $\lambda > \Delta$. Whereas, if $\lambda_{\sigma} = \lambda$, i.e, spin-independent then the Hamiltonian breaks time-reversal symmetry and the ground state represents a QH state when $\lambda>\Delta$. In what follows, it is not required to have a finite $\Delta$ but typically it helps in numerical stability. In passing we note that, a time-reversal symmetry breaking $\lambda$ can be introduced through a circularly polarized irradiation on the sample~\cite{Ezawa2011a,graFl1,graFl2}, which can provide a way for fine tuning the parameter. 

 \begin{figure}[t]
 	\centering
 	\includegraphics[width=0.42\textwidth]{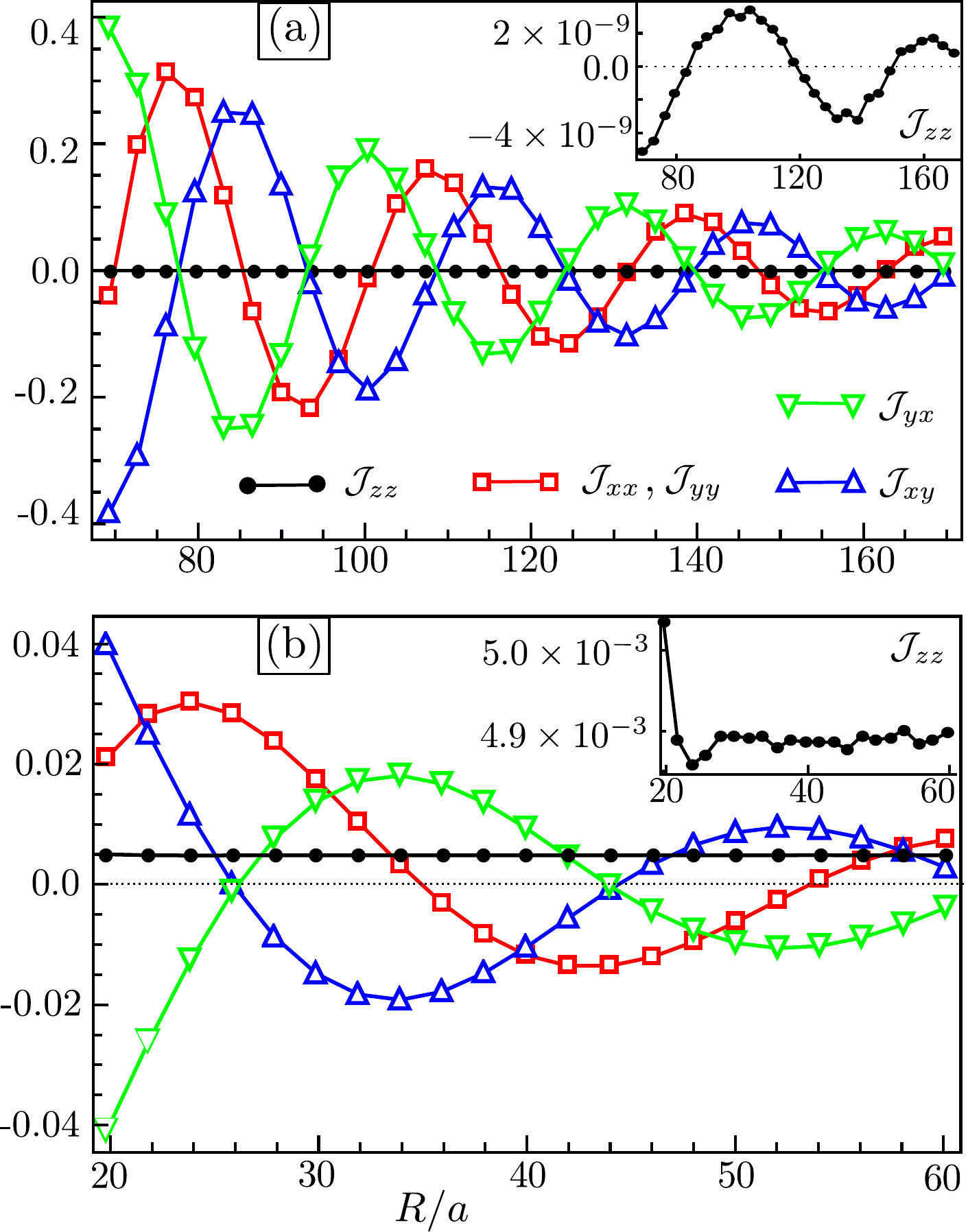}\\
 	\caption{(a) For an infinite nano-ribbon of QSH state, the RKKY interaction between two spins put on a edge is significantly small when the spins' moments are aligned in up/down direction, resulting from the chiral nature of the edges.  (b) In contrast, for a finite geometry, even when the spins' moments are along up/sown direction, the RKKY interaction between them mediated by the edge states is not small. For numerical simulation in (a), a zigzag nano-ribbon of width $16$ sites has been used whereas in (b), a system size of $N_x\times N_y=80\times$16 sites has been use with the impurity spins on the longer zigzag edge. Other parameters used are $\lambda = 0.5$, $\Delta = 0.1$, $t=1.0$.}\label{fig:rkkyfin}
 \end{figure}

The result of preceding section, i.e, vanishing $\mathcal{J}_{zz}$ correlation for an infinite QSH edge, can be verified using an infinite nano-ribbon geometry, shown in Fig.~\ref{fig:rkkyfin}. All other terms in the correlation matrix is generally non-zero, including the off-diagonal elements, resulting in Dzyaloshinskii-Moria interaction among the spins~\cite{RKKYQSH1}. 

\begin{figure}[t]
	\centering
	\vspace{0.2cm}
	\includegraphics[width=0.25\textwidth]{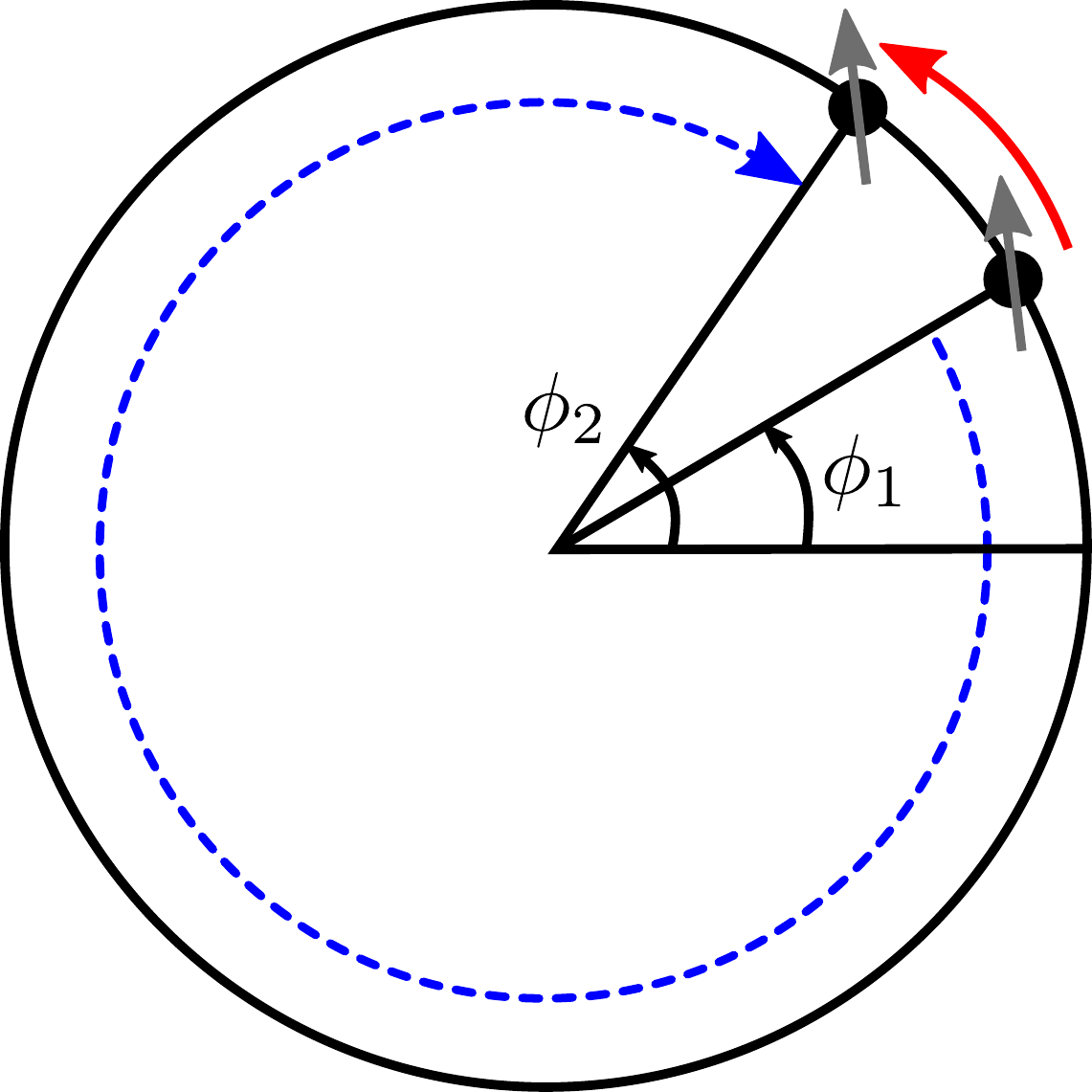}\\
	\caption{For a disk geometry of the QSH system, the edge states run along the edge of the disk with impurity spins put in positions $\phi_1$ and $\phi_2$. The spins can be connected via two paths as shown, which makes all relevant couplings between the spins non-vanishing.}\label{fig:ring}
\end{figure}

Instead of an infinite  nano-ribbon, for a finite geometry, using the Green's function $G(\omega) = [\omega -H + i0^+]^{-1}$ in real space, the RKKY interaction can be computed as second order perturbation, Eq.~(\ref{eq:rkky}). The impurity spins can be taken into account within $H$ using the Kondo coupling between the localized spins, $\mathbf{S_1},\mathbf{S}_2$, put at site $i,j$, and the delocalized electrons given by 
\begin{align}\label{eq:H'}
H' = -\frac J2\sum_{\sigma\sigma'} c^{\dagger}_{i\sigma}S_{1,\sigma\sigma'}c_{i\sigma'} + c^{\dagger}_{j\sigma}S_{2,\sigma\sigma'}c_{j\sigma'}.
\end{align}
The RKKY interaction can also be obtained using an exact diagonalization method in a finite geometry~\cite{graBS}. Despite the fact than the exact diagonalization is numerically less expensive, we use the Green's function method as it would provide more flexibility, especially for an open system as we shall discuss later. The exact diagonalization result matches with the second order perturbation in the limit when $J$ is sufficiently small~\cite{graSat,graBS}.

The resulting diagonal terms of interaction matrix among the spins put on the edge of a finite QSH geometry is plotted in Fig.~\ref{fig:rkkyfin}, which is markedly different than predicted through Eq.~(\ref{eq:Gupdn}), as the interaction among the spins is non-zero even when the spins are both up or down, i.e, $\mathcal{J}_{zz} \neq 0$. As the chiral nature of the modes are still present, this breakdown from the previous result is unexpected. 

To explore the reason for such we take a simple geometry of a disk, where the chiral modes can run through its perimeter. The Hamiltonian of the 1D edge modes is then $H_{ch} = \frac{v_F\sigma_z}{R}(-i\partial_{\phi}+\frac12)$, where $R$ is the radius of the disk and $\phi$ is the azimuthal angle. Angular momentum $-i\partial_{\phi}$ is quantized with energy eigenvalues $E_{l} = \sigma \frac{v_F}{R} \left(l+\frac12\right)$~(see Appendix \ref{app:II}). The Greens function becomes
\begin{align}
G_{\sigma\sigma}(\phi,\phi';\omega)= \sum_l \frac{e^{i(\phi-\phi') l}/2\pi}{R\omega + iR\eta - \sigma v_F (l+1/2) },
\end{align}
where $\sigma=\pm1$ is for up and down spin blocks respectively. Note that, it is not possible, in general, to convert this summation into an integral form as the integrand changes swiftly from $l$ to $l+1$, unless the phase $\phi$ is very small. In contrast to Eq.~(\ref{eq:Gupdn}), this Green's function, in the limit when $ \eta< v_F/2\pi R$, can now connect between arbitrary points on the circle (i.e, $|G_{\sigma\sigma}(\Delta\phi,\omega)|$ is non-zero for all $\Delta\phi = \phi-\phi'$) for both the spin modes~(see Appendix \ref{app:III}). This essentially captures that in a finite geometry the interaction among the spins is possible through two possible paths (Fig.~\ref{fig:ring}). In the second order process, Eq.~(\ref{eq:corr}), it turns out that the integrand of Eq.~(\ref{eq:rkky}) for the $\mathcal{J}_{zz}$ component becomes independent of the difference between the azimuthal angles, $\Delta \phi$, giving rise to distant independent coupling between the spins~\cite{analyze}. Such distance-independent nature of correlation is true only for the $\mathcal{J}_{zz}$ coupling, whereas other correlation matrix elements remain function of $\Delta \phi$. An analytical explanation is given in Appendix \ref{app:IV}. The results of the Fig.~\ref{fig:rkkyfin} in light of this discussion is one of our main result.

\begin{figure}[t]
	\centering
	\includegraphics[width=0.46\textwidth]{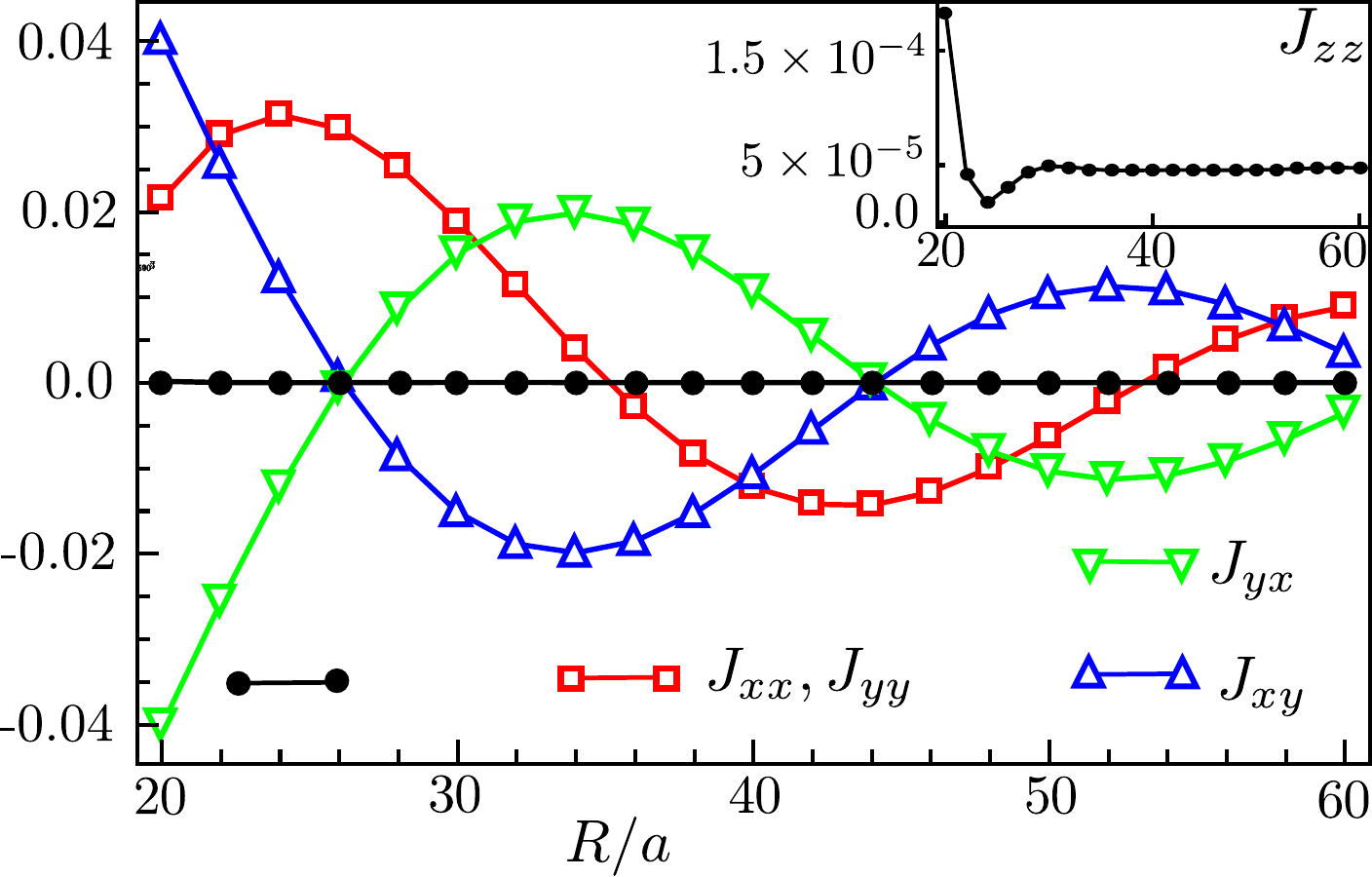}\\
	\caption{RKKY coupling strengths $\mathcal{J}_{ij}$ for a QSH system that is connected to a lead on one side. The out-of-plane component, $\mathcal{J}_{zz}$ vanishes as the edge states suffer from the decoherence introduced by the lead. For numerical simulation $\pi\rho t_0^2 = 0.5$ (in the unit of $t$) has been used to couple to the right side of the system, where $t_0$ is the coupling amplitude between the system and the lead. Other parameters are of that of Fig.~\ref{fig:rkkyfin}(b). The nature of such reduction with the lead's density of states has been discussed further in Fig.~\ref{fig:rho}.}\label{fig:lead}
\end{figure}

The inversion broken nature of the QSH states also results in finite non-collinear Dzyaloshinskii-Moria interaction among the impurity spins, which is present in either finite or infinite geometry. For the QH edge, a similar observation of QSH is made, that, instead of vanishing correlation matrix $\mathcal{J}_{ij}$, one observe non-vanishing values for all of elements. In the simple model of QH,  Eq.~(\ref{eq:H0}), the diagonal elements of the correlation matrix become independent of the position between the spins whereas the off-diagonal elements remain zero.

The preceding discussion is strictly true if the perimeter of the finite QSH geometry is not larger than the mean free path of the electrons at the edge states (see Appendix \ref{app:III} for details). As the QSH edge prohibit back scattering, one expects a large mean free path of order few hundreds of nanometers~\cite{GAAS}. The consideration of a finite mean-free path would affect in further decay of all the elements of the spin-spin correlation matrix $\mathcal{J}$.

\emph{Non-local control using leads.}--- Given the different nature of interaction among spins in an infinite and finite geometry, one natural question is whether such difference can be engineered without actually altering the geometry. Essentially prohibiting the edge states from fully traversing the perimeter should mimic the behavior of the infinite geometry. Such a situation can be engineered using a lead attached to the edge far from the two impurities. Then for a sufficiently strong system-lead coupling, all the edge states will go inside the lead and the coherence will be lost.

We proceed to treat the system with the lead attached by considering a self-energy contribution to the Green's function of the system 
\begin{align}\label{eq:GF}
G(\omega)^{-1} = \omega -H - i\Sigma(\omega),
\end{align}
where $H$ is the Hamiltonian of the system without the lead and $\Sigma(\omega) = -i K^{\dagger}g(\omega)K$ with the system-lead coupling matrix $K$ and the lead green function $g(\omega)$. In the system of our consideration, the QSH edges, live within a bulk gap and it is expected that most of the contribution in the RKKY interaction comes from states near the Fermi energy. So, without loss of generality, it is sufficient to take the approximation that the lead green function is independent of energy and we simply write $g(\omega) \approx i\pi\rho$, where $\rho$ is the (energy independent) density of states of the lead at the contact. $\rho$ determines our strength of system-lead coupling.

The effective Green's function, Eq.~(\ref{eq:GF}), is now written is real space, where the impurity spins can be taken into account within $H$ as in Eq.~(\ref{eq:H'}). The effective Green's function can be used back in Eq.~(\ref{eq:rkky}) to compute the effective interaction between the spins mediated by the underlying system.

The energy integral in Eq.~(\ref{eq:rkky}) should in principle run for the full bandwidth below the Fermi energy, which we have used in our numerical simulation. But in practical systems, a finite mean free path (i.e, a finite lifetime of the states) of the electron would result in a much smaller effective range of the integral if the spins are located far enough. Moreover, the bulk states of QSH are not protected against back scattering (i.e, not chiral), so one expects a much shorter mean free path of the bulk states compared to the edge states. For simplicity, one can simply restrict the energy integral in Eq.~(\ref{eq:rkky}) within the gap in the bulk spectrum.

The result of adding a lead is summarized in Fig. \ref{fig:lead}, where, as expected, we observe a sharp drop the $zz$ component of the interaction in presence of the lead, whereas other components are effectively the same as Fig.~\ref{fig:rkkyfin}(b). The $\mathcal{J}_{zz}$ drops as an exponential function of the lead density of state, but the drop becomes slower after a threshold value of $\rho$ is reached. This threshold value of $\rho$ also depends on the details of the system-lead coupling, such as the area of the system that is connected to the lead. In our simulation, we have attached the lead at one of the side of the system, which has $80\times16$ sites. The selective action of the lead to the Ising interaction is a direct demonstration of the helical nature of the edge states. This can provide a way to identify helical nature of edge states as well as can be used as effective spin-control in spintronic setup. This is one of our main results.

\begin{figure}[t]
	\centering
	\includegraphics[width=0.46\textwidth]{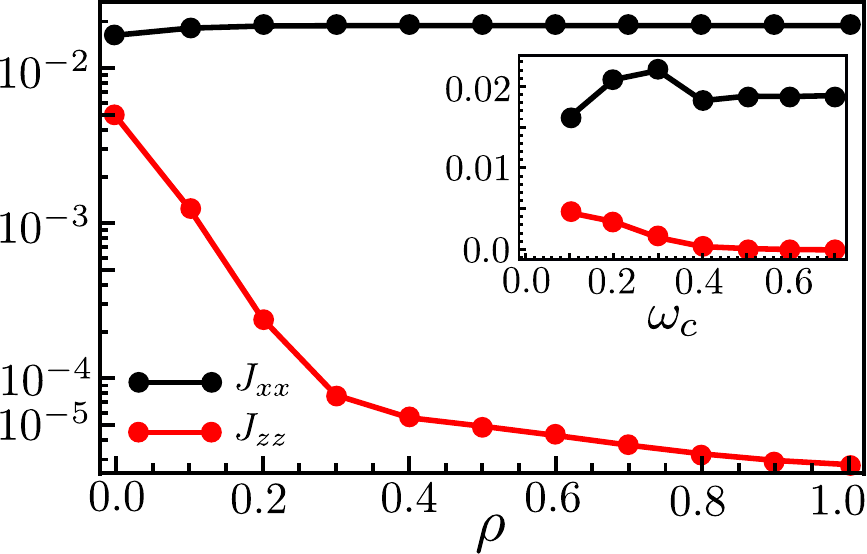}\\
	\caption{ With increasing density of states ($\rho$) of the lead, the out-of-plane correlation $\mathcal{J}_{zz}$ vanishes whereas other coupling terms remain virtually unaffected. Inset: for a $\rho$ which is 0.5 for $0>\omega>\omega_c$, the $\mathcal{J}_{zz}$ rapidly decreases as a function of $\omega_c$ until it reaches the bulk gap, which is, in our simulation 0.4. The two impurities are at $(8, N_y)$ and $(30, N_y)$ and all other parameters are same as Fig. \ref{fig:rkkyfin}.}\label{fig:rho}
\end{figure}

In realistic setup, the lead's density of state will be dependent on the energy, but our main finding should remain intact. In fact, it is sufficient to consider the lead as a quantum dot with a broadening of its level of the order of the spin-orbit gap ($\delta = \lambda - \Delta$) of the system, which is typically of the order of a few milli-electron volts. We further show in the Fig.~\ref{fig:rho} that as long as the $\rho(\omega)$ is non-zero for the range of $\omega$ within the bulk gap of the QSH system, the $\mathcal{J}_{zz}$ correlation remains vanishingly small. Whereas, as soon as $\rho(\omega)$ is zero for a range of $\omega$ where edge states exist, $\mathcal{J}_{zz}$ acquires a finite value (see Fig.~\ref{fig:rho} inset). This provides a concrete way to control the interaction among the spins: for turning the interaction on or off it is sufficient to either change the quantum dot's (which now act as a lead) band gap through a gate bias or the system to lead coupling. For a QH system such arrangement can control the full spin-spin correlation matrix.

Although unrealistic, the same result can be recovered using a finite $i\eta$ ($\eta>0$) added in the right hand side of the Eq.~(\ref{eq:GF}) instead of the lead, which would basically add a finite lifetime to \textit{all} the eigenstates. As the states near the Fermi energy are moving with the Fermi velocity, the coherence is present only for a given length of their path (i.e, mean free path) given by $\sim v_F/\eta $. If the perimeter is larger than the finite length, then Ising interaction would inevitably vanish. But a finite $\eta$ will effect also other possible interactions among the spins. When $\eta\rightarrow0$, one recovers the exact diagonalization result of Fig.~\ref{fig:rkkyfin}(a).

As mentioned earlier, the coupling $\lambda$ can be introduced and controlled in a system without any intrinsic spin-orbit interaction (such as graphene) using a circularly polarized light. Although the viability of application of such system is still under investigation, application of terahertz radiation in spintronics application is viable~\cite{RKKYTerahz,tera}. Such an arrangement provides a further way for controlling the system parameter $\lambda$. 

\emph{Discussion.}--- 
The effective system size we have taken is of the order of a few nanometers (about 100 lattice spacings of typical solid state systems). Our other scales, essentially the spin-orbit coupling $\lambda$ is taken to be large enough (for the parameters, please see the figure caption of Fig.~\ref{fig:rkkyfin}) for the benefit of numerical simulation. A larger spin-orbit coupling provides a larger band-gap in the bulk, although a larger $\lambda$ results in smaller spin-spin correlation~(see Appendix \ref{app:V}). In realistic systems, even if the SO coupling is smaller, the system size can be much larger, so that the RKKY mediated by the bulk states can still be neglected. To clearly observe the physics we propose, one needs to have a system with the size of perimeter much larger than the mean free path of the bulk states, but shorter than the mean free path of the edge states. Whereas the system-lead coupling can not be controlled effectively, the density of states of the lead (at the Fermi energy of the system) can be controlled electrically using a semiconducting lead with controllable band gap or a quantum dot using another gate. 

Interaction in the one-dimensional channels of a QH or QSH edge can fractionalize the modes and in general more than one propagating modes will be present and most of the conclusions of an infinite edge follows similarly~\cite{RKKYQSH2}. A general formalism for treating a finite system is left for futures study but we expect the physics behind Fig.~\ref{fig:ring} to remain intact, giving rise to breakdown of earlier results. In passing we note that, interestingly, distant-independent and non-oscillating RKKY interaction has also been reported in interacting graphene system~\cite{graU}, although the mechanism is different. With interaction graphene edges becomes spin-polarized rendering anti-ferromagnetic orientation costly irrespective of distance.

In summary, we theoretically demonstrate how the effective interaction among spins on a QSH or QH edge differ in nature in a finite system compared to an infinite edge. The difference in nature can be observed using a lead attached to the system with controllable density of states of the lead. This provides, at one hand a way to identify helical edges of a system as well as a truly non-local way to control the interaction among the spins, which might be useful in quantum information and spintronic applications.

A.K. would like to thank useful communication with H.A. Fertig (IU Bloomington), S. Satpathy (Univ. Missouri), M. Sherafati (Truman State Univ.) and S. Mukhopadhyay (IIT Kanpur) at various parts of the work.


\section*{Appendix}
\section{Results of square lattice}\label{app:I}
\begin{figure}[t]
	\begin{center}
		\includegraphics[width=0.45\textwidth]{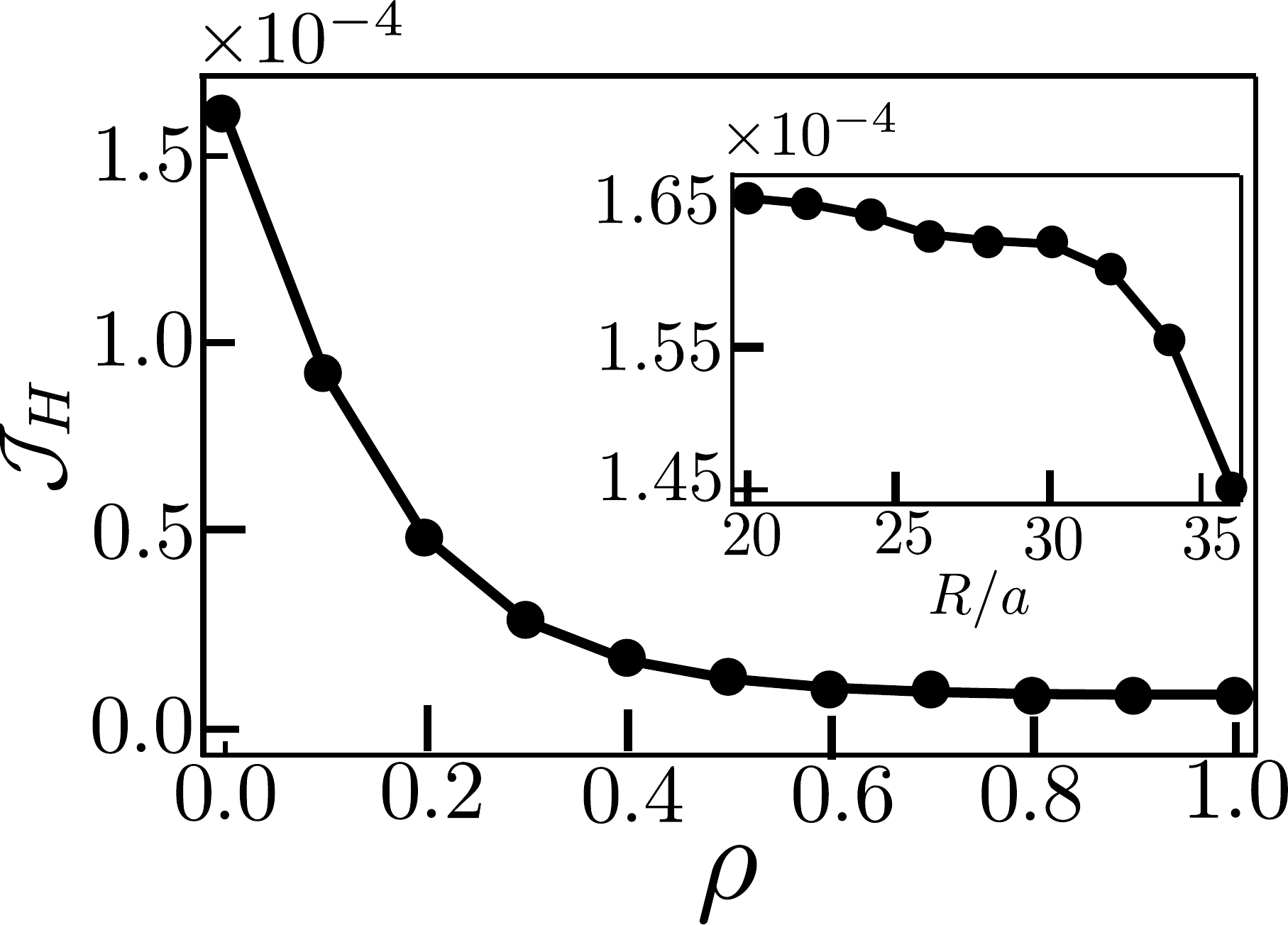}
		\caption{Spin correlation of the edge of the QH system in a square lattice. The Heisenberg type coupling $\mathcal{H}$ (in the unit of $J^2$) is almost independent of distance (and anti-ferromagnetic) as shown in the inset. The coupling can be controlled using a lead, as discussed in the main text, with density of state $\rho$. System size $40\time 20$, $A=B=1$, $M=-0.3$, $\eta=0.0025$. For main plot the two impurities are at $(8, N_y)$ and $(30, N_y)$ and the lead is attached throughout the right edge of the system. For inset the first impurity is fixed at $(8, N_y)$ and all other parameters are same as the main plot.  }\label{fig:square}
	\end{center}
\end{figure}

We consider Bernevig-Hughes-Zhang model for quantum Hall state in 2D square lattice:
$H(\textbf{k}) = A\tau_{x}\sin k_x + A\tau_y\sin k_y + B \tau_z( 2 + M - \cos k_x -\cos k_y)$,  $\sigma$'s are Pauli matrices of pseudo spin and with lattice constant unity.  This model describes topological insulator phase for $-2<M<0$ and $-4<M<-2$ with Chern number $\mp1$ respectively.  We consider here $A=B=1$, $M=-0.3$ and system size $N_x\times N_y = 40\times 20$. We summarize the results of RKKY interaction when two spins are put on the longer edge of the system in Fig.~\ref{fig:square}. First, similar to the hexagonal-lattice model described in the main text, we obtain a anti-ferromagnetic couplings for the diagonal terms of the correlation matrix, which depends weekly on the distance between the spins. The coupling can be tuned using a load far from the impurities.

\section{Effective energy dispersion of quantum Hall finite-edge}\label{app:II}
For $-2<M<0$ around $\Gamma =(0,0)$ point the effective Hamiltonian for a quantum Hall system is given by
\begin{align}
	H= v_{F}(k_x \sigma_x + k_y\sigma_y + m\sigma_z),
\end{align} 
here $\sigma$'s are the Pauli's matrices describing the pseudo-spin degree of freedom. $m$ opens gap between bulk states, but there are gapless topological edge states.  The total angular momentum operator $J_z = L_z + \frac{1}{2}\sigma_z$ commutes with the above Hamiltonian, so they both have a common complete set of eigenstates. Our aim is to find out the effective dispersion of the edge states of a QH state, for simplicity, in a disk geometry. So now onward we will use polar co-ordinates: $x=r\cos{\phi}$ , $y=r\sin{\phi}$, $r=\sqrt{x^2 + y^2}$, $\phi = \arctan{(y/x)}$. Eigenstates of $J_z$ will be of the form
\begin{align}
	\psi_{n^{\prime}}(r,\phi)=\begin{pmatrix}
		e^{i (n^{\prime}-1/2)\phi}f_{n^{\prime}}(\phi)\\
		e^{i (n^{\prime}+1/2)\phi}g_{n^{\prime}}(\phi)
	\end{pmatrix}.
\end{align}
The wave function should be single valued i.e. $\psi(r,\phi + 2\pi)=\psi(r,\phi)$, which immediately implies that $n^{\prime}-1/2 = n$ should be an integer. So final form of the eigen functions will be
\begin{align}
	\psi_n(r,\phi)=\begin{pmatrix}
		e^{in\phi}f_{n}(r)\\
		e^{i(n+1)\phi}g_{n}(r)
	\end{pmatrix}.\label{eq:ES}
\end{align}
Now we consider the following boundary value problem. In the region $I$, defined as $r<R$, we consider the mass $m_I = m$ where $m>0$ and in region $II$, where $r>R$ we have $m_{II} = -m$. The change of Chern number across the boundary is $\pm 1$ and consequently there will an edge state along the boundary that decays exponentially to both $r>R$ and $r<R$ regions. By matching the wave functions at the  boundary of the disk, we find the energy dispersion for QH edge. The Hamiltonian is the polar coordinate is
\begin{align}
	H=&v_F(m\sigma_z - i(\sigma_x\cos{\phi}+ \sigma_y \sin{\phi})\partial_{r}) \nonumber\\
	& + i(\sigma_x\sin{\phi} - \sigma_y \cos{\phi})r^{-1}\partial_{\phi})).
\end{align}
Using the form of the eigenstate Eq.~(\ref{eq:ES}), we get
\begin{align}
	\left(\partial_r + \frac{n+1}{r}\right)g_n(r) = ((\epsilon_n/v_F) + m) f_n(r)\nonumber\\
	\left(-\partial_r + \frac{n}{r}\right)f_n(r) = ((\epsilon_n/v_F) - m) g_n(r),
\end{align}
where $\epsilon_n$ is the eigenenergy. 
The physical solution of the equations for $0<r<R$  is $f_n(r) = a_n \mathcal{J}_{n}(\tilde{\epsilon}_n r)$. and $g_n(r) = a_n \frac{\tilde{\epsilon}_n}{(\epsilon_n/v_F) - m}\mathcal{J}_{n + 1}(\tilde{\epsilon}_n r)$, here where $\tilde{\epsilon}_n = \sqrt{(\epsilon_n/v_F)^2 - m^2}$ and  $\mathcal{J}_{n}(.)$ is the Bessel function of first kind and $a_n$ is a constant. So the wave function for region $I$ is
\begin{align}
	\psi^{I}({r,\phi})=\sum_{n}\frac{a_n}{\sqrt{2}} \begin{pmatrix}
		e^{in\phi}\mathcal{J}_{n}( \tilde{\epsilon}_n r)\\
		i \frac{\tilde{\epsilon}_n}{(\epsilon_n/v_F)-m}e^{i(n+1)\phi}\mathcal{J}_{n+1}( \tilde{\epsilon}_n r).\label{eq:WF1}
	\end{pmatrix}
\end{align}
Similarly for region $II$ the wave function will be
\begin{align}
	\psi^{II}({r,\phi})=\sum_{n}\frac{b_n}{\sqrt{2}}\begin{pmatrix}
		e^{in\phi}\mathcal{H}_{n}^{(1)}( \tilde{\epsilon}_n r)\\
		i  \frac{\tilde{\epsilon}_n}{(\epsilon_n/v_F)+m}e^{i(n+1)\phi}\mathcal{H}_{n+1}^{(1)}( \tilde{\epsilon}_n r),\label{eq:WF2}
	\end{pmatrix}
\end{align}
where $\mathcal{H}_{n+1}^{(1)}(.)$ is the Hankel function of first kind. At the boundary  of the disk
\begin{align}
	\psi^{I}({R,\phi}) =\psi^{II}({R,\phi}).\nonumber
\end{align}
Using the wave functions we get
\begin{align}
	\epsilon_n = mv_F\frac{\mathcal{J}_{n}(\tilde{\epsilon}_n R)\mathcal{H}_{n+1}^{(1)}(\tilde{\epsilon}_n R) + \mathcal{H}_{n}^{(1)}(\tilde{\epsilon}_n R)\mathcal{J}_{n+1}(\tilde{\epsilon}_n R)}{\mathcal{J}_{n}(\tilde{\epsilon}_n R)\mathcal{H}_{n+1}^{(1)}(\tilde{\epsilon}_n R) - \mathcal{H}_{n}^{(1)}(\tilde{\epsilon}_n R)\mathcal{J}_{n+1}(\tilde{\epsilon}_n R)}.\nonumber
\end{align}
using the identity
\begin{align}
	\mathcal{J}_{n}(x)\mathcal{H}_{n+1}^{(1)}(x) - \mathcal{H}_{n}^{(1)}(x)\mathcal{J}_{n+1}(x) = -2 i/(\pi x),\nonumber
\end{align}
the equation becomes
\begin{align}
	\epsilon_n = mv_F(1 + i\pi (\tilde{\epsilon}_n R) \mathcal{H}_{n}^{(1)}(\tilde{\epsilon}_n R)\mathcal{J}_{n+1}(\tilde{\epsilon}_n R)).\nonumber
\end{align}
Now,  $\mathcal{J}_{n}(i z))=e^{i(n+1)\pi/2} \mathcal{I}_{n}(z)$ and  $\mathcal{H}_{n}^{(1)}(i z))=e^{-i(n+1)\pi/2} \frac{2}{\pi}\mathcal{K}_{n}(z)$, where $ \mathcal{I}_{n}(z)$ and $ \mathcal{K}_{n}(z)$ are modified Bessel functions of first and second kind respectively. In the limit $M\gg \epsilon_n$, the asymptotic forms of these functions are $\mathcal{I}_{n}(z)\sim\frac{e^{z}}{\sqrt{2\pi z}}(1-\frac{4n^2-1}{8z})$, 
$\mathcal{K}_{n}(z)\sim e^{-z}\sqrt{\frac{\pi}{2 z}}(1+\frac{4n^2-1}{8z})$.
Using these we get the energy eigenvalues
\begin{align}
	\epsilon_n = \frac{v_F}{R}\left(n +\frac{1}{2}\right).\label{eq:EV}
\end{align}

\section{Chirality of the Green's function}\label{app:III}
For a simplified description of the edge state, we consider the wavefunction Eq.~(\ref{eq:WF1}) at $r=R$ along with the solution Eq.~(\ref{eq:EV}) and the normalization $\psi_{n}^{\dagger}(R,\phi)\psi_n(R,\phi)=1/(2\pi R)$, giving
$$a_{n}\sqrt{2\pi R}=1/\sqrt{|\mathcal{J}_{n}(\tilde{\epsilon}_n R)|^2 + \frac{|\tilde{\epsilon}_n|^2}{((\epsilon_n/v_F)-M)^2}|\mathcal{J}_{n+1}(\tilde{\epsilon}_n R)|^2}.$$ The Green's function is of the form
\begin{align}
	G(\phi,\phi^{\prime},\omega + i\eta)
	\equiv&\sum_{n}\frac{\alpha}{\omega'+ i\eta'-n^{\prime}}\begin{pmatrix}
		G_{AA} & G_{AB}\\
		G_{BA} & G_{BB}
	\end{pmatrix},\label{eq:GF}
\end{align}
where $\alpha=R/v_F$, $\omega'=\omega\alpha$, $\eta'=\eta\alpha$, $n^{\prime}= n + \frac{1}{2}$ and
\begin{align}
	G_{AA}=&\sum_{n}\frac{\alpha a_{n}^{2}e^{-i n(\phi-\phi^{\prime})}}{\omega'+ i\eta'-n^{\prime}}\mathcal{J}_{n}(\tilde{\epsilon}_n R)\mathcal{J}_{n}^{*}(\tilde{\epsilon}_n R),\nonumber\\
	G_{AB}=&\sum_{n}\frac{-i\alpha a_{n}^{2}e^{-i(n+1)\phi}e^{i n\phi^{\prime}}}{\omega'+ i\eta'-n^{\prime}}\mathcal{J}_{n}(\tilde{\epsilon}_n R)\mathcal{J}_{n+1}^{*}(\tilde{\epsilon}_n R),\nonumber\\
	G_{BA}=&\sum_{n}\frac{i\alpha a_{n}^{2}e^{-i n\phi}e^{i (n+1)\phi^{\prime}}}{\omega'+ i\eta'-n^{\prime}}\mathcal{J}_{n+1}(\tilde{\epsilon}_n R)\mathcal{J}_{n}^{*}(\tilde{\epsilon}_n R),\nonumber\\
	G_{BB}=&\sum_{n}\frac{\alpha a_{n}^{2}e^{-i (n+1)(\phi-\phi^{\prime})}}{\omega'+ i\eta'-n^{\prime}}\mathcal{J}_{n+1}(\tilde{\epsilon}_n R)\mathcal{J}_{n+1}^{*}(\tilde{\epsilon}_n R).\nonumber
\end{align}
Upto this we have considered only up-spin. For a QH system, the down spin wave function and the Green's function will be identical to up spins' whereas for QSH system, for the down spin, we need to replace $v_F$ by $-v_F$. We consider general positions of the impurity spin $\textbf{S}_{1}$ is at $(R,\phi_1)$ and that of impurity spin $\textbf{S}_{2}$ at $(R,\phi_2)$ and we write $\phi_1 - \phi_2 =\phi_{12}$. Both $\phi_1$ and $\phi_2$ can take values between $0$ to $2\pi$. For studying the RKKY interaction between these impurities in a QSH system we rewrute the Green's functions as
\begin{align}
	G_{AA}^{\sigma}(\phi_{12}, \omega + i\eta)
	=&\sum_{n}\frac{\alpha C_{AA}^{n}e^{-i n\phi_{12}}}{\omega'+ i\eta'-\sigma n^{\prime}},\label{eq:analytic}\\
	G_{AB}^{\sigma}(\phi_{12}, \omega + i\eta)
	=&\sum_{n}\frac{-i\alpha C_{AB}^{n}e^{-i (n+1)\phi_{1}}e^{i n \phi_{2}}}{\omega'+ i\eta'-\sigma n^{\prime}},\nonumber\\
	G_{BA}^{\sigma}(\phi_{12}, \omega + i\eta)
	=&\sum_{n}\frac{i\alpha C_{BA}^{n}e^{-i n\phi_{1}}e^{i (n+1) \phi_{2}}}{\omega'+ i\eta'-\sigma n^{\prime}},\nonumber\\
	G_{BB}^{\sigma}(\phi_{12}, \omega + i\eta)
	=&\sum_{n}\frac{\alpha C_{BB}^{n}e^{-i (n+1)\phi_{12}}}{\omega'+ i\eta'-\sigma n^{\prime}},\nonumber
\end{align}
where $\sigma = \pm 1$ for up and down spins respectively, $C_{AA}^{n} =  a_{n}^{2}|\mathcal{J}_{n}(\tilde{\epsilon}_n R)|^2$, $C_{AB}^{n} =  a_{n}^{2} \mathcal{J}_{n+1}^{*}(\tilde{\epsilon}_n R) \mathcal{J}_{n}(\tilde{\epsilon}_n R)$, $C_{BA}^{n}=  a_{n}^{2} \mathcal{J}_{n}^{*}(\tilde{\epsilon}_n R) \mathcal{J}_{n+1}(\tilde{\epsilon}_n R)$ and $C_{BB}^{n}=  a_{n}^{2} |\mathcal{J}_{n+1}(\tilde{\epsilon}_n R)|^2$. These forms will be used in the next section.

Before we proceed to compute the RKKY interaction, we make some comments about chirality of the Green's function. By observing that $C_{AA}^{n}$ does not depend on $n$, one can write the equivalent Green's function in the sublattice sector $AA$ as
\begin{align}
	G^{\sigma}_{AA}(\phi,\omega)	=& \sum_n \frac{1}{2\pi v_{F}}\frac{ e^{i\phi n}}{\omega' + i\eta' - \sigma (n+\frac{1}{2}) }.\label{eq:Gapp}
\end{align}
Note that, it is not possible, in general, to convert this summation into an integral form as the integrand changes swiftly from $n$ to $n+1$, unless the phase $\phi$ is very small. The amplitude $|G^{\sigma}_{AA}(\phi,\omega)|$ dictates how the propagating modes can connect points which are at position $(\phi_0,R)$ and $(\phi_0 + \phi,R)$. We below describe the two different physical regimes of interest for values of $\eta$:
\begin{figure*}[t]
	\begin{center}
		\includegraphics[width=0.7\textwidth]{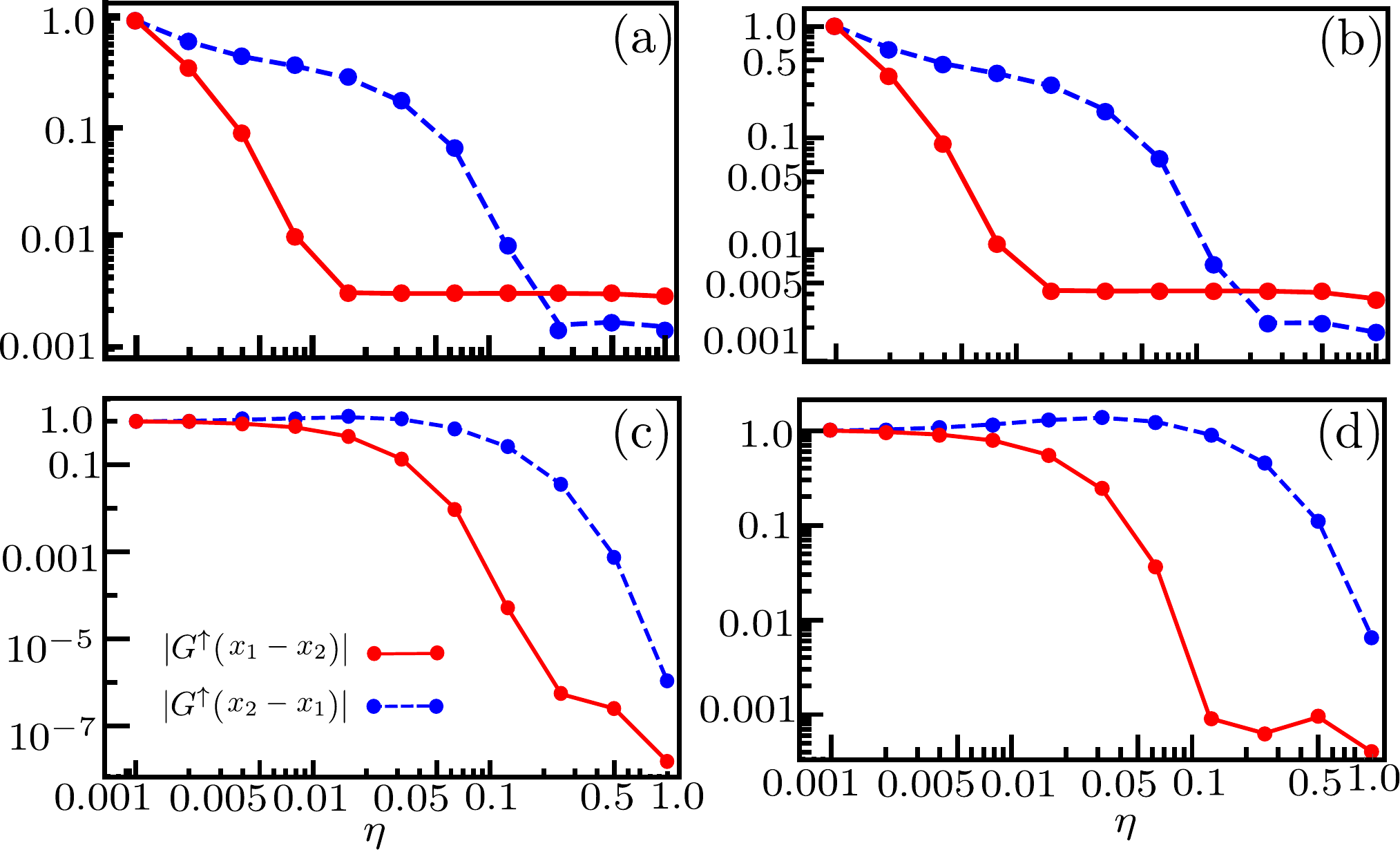}
		\caption{Due to the chiral nature of the edge states, two points ($x_1,x_2$) on the edge of the quantum Hall system are connected differently depending on the sign of $(x_1-x_2)$. The difference is negligible if the mean free path is much larger (i.e, $\eta$ is smaller) than the perimeter of the system.  To show this, we plot the absolute values of Green's functions normalized by their values at a very small $\eta= 0.001$ for a given $\omega = -0.015$. For the approximated Green's function, Eq.~(\ref{eq:Gapp}) (shown in (a)), for the analytical Green's function,  Eq.~(\ref{eq:analytic})	 (shown in (b)), for the lattice model in the square lattice (shown in (c)) as well as the model in the hexagonal lattice (shown in (d)), we observe the similar behavior that the relative difference between $|G^{\uparrow}(x_1-x_2)|$ and $|G^{\uparrow}(-x_1+x_2)|$ becomes more apparent with increasing $\eta$ until a very large $\eta$ is reached. The physics of the paper is valid when the relative difference is maximum.
For (a) and (b) we use $R=100$, $\phi = 0.1\pi$, $v_{F} = 1.0$, and for (b), $m = 10$. Numerical results for finite geometry QH system in (c) we obtain  with system size $N_x\times N_y = 30\times 20$, $A=B=1$, $M=-0.8$, and the impurities are at $(4,N_y)$ and $(20,N_y)$. For (d) using Haldane-Kane-Mele Model model we use system size $N_x\times N_y = 60\times 20$, $\lambda= 0.5$,$lE_z = 0.1$, $t=1.0$, and the impurities are at $(4,N_y)$ and $(20,N_y)$. }\label{fig:Green}
	\end{center}
\end{figure*}
\begin{enumerate}
	\item When the mean free path is smaller than the perimeter of the disk, $\lambda \ll 2\pi R$. As the mean free path $\lambda \sim v_F/\eta$, this is similar condition as $\eta R/v_F \gg 1/2\pi$. In this case it is expected that the helical modes can not fully traverse through the perimeter without decoherence. For a fixed value of $\omega$ it is easy to verify that $G_{AA}^{\sigma}(\phi,\omega)$ connects chirality and this we show in the Fig.~\ref{fig:Green}.
	\item When the mean free path is larger than the perimeter of the disk, $\lambda > 2\pi R$, which is similar condition as $\eta R/v_F < 1/2\pi$. In this case it is expected that the helical modes can fully traverse through the perimeter only a few time without decoherence (but it decays as it traverses). For a fixed value of $\omega$ it is easy to verify that $|G^{\sigma}_{AA}(\phi,\omega)|$ is not fully chiral as a result of this and this we also depict this in the Fig.~\ref{fig:Green}.
\end{enumerate}
In regime 1, the system, although in a ring, behaves just like an infinite 1D chain, the resulting correlation behaves similar to that of an infinite quantum spin-Hall edge. The work of this paper falls in regime 2, where the states can traverse the full perimeter.

In a finite geometry only a finite number of edge states can contribute to spin-spin correlation. This number is controlled by $R$ in this model. An upper bound in $n$ is also important to impose as the Green's function is a slowly varying function in $\omega$. Finally, this model misses any other effects that might be present in a rectangular geometry, resulting from quantum oscillations and corner effects. Further, the sub-lattice structure of the hexagonal lattice provides a wave function with features that the simple model will fail to capture. Only the qualitative picture we expect to remain intact in an arbitrary geometry. We also show in the figure how the basis understanding of the discussion remain true for various models as well as by using the Green's function of Eq.~(\ref{eq:GF}).

\section{RKKY interaction}\label{app:IV}
Consider the general form of RKKY interaction
\begin{align}
	&H_{\text{RKKY}}=\nonumber\\
	& -\frac{J^2}{\pi} {\rm Im}\int_{-\infty}^{E_F} d\omega \text{Tr}[(\mathbf{S}_1.\mathbf{\sigma})G_{\mu\nu}(\phi_{12},\omega)(\mathbf{S}_2.\mathbf{\sigma})G_{\mu\nu}(-\phi_{12},\omega)]\nonumber\\
	& \equiv  \mathcal{J}_{I}S_{1z}S_{2z} + \mathcal{J}_{H}(S_{1x}S_{2x} + S_{1y}S_{2y})+ \mathcal{J}_{DM}(\textbf{S}_{1}\times\textbf{S}_{2})_z\nonumber,
\end{align}
here $\mu, \nu = A/B$, which gives various RKKY coupling as
\begin{widetext}
	\begin{align}
		\mathcal{J}_{I}=\frac{-J^2}{\pi} {\rm Im}\int_{-\infty}^{E_F} d\omega [G_{\mu\nu}^{+}(\phi, \omega)  G_{\mu\nu}^{+}(-\phi, \omega) +  G_{\mu\nu}^{-}(\phi, \omega)  G_{\mu\nu}^{-}(-\phi, \omega)],\\
		\mathcal{J}_{H}=\frac{-J^2}{\pi} {\rm Im}\int_{-\infty}^{E_F} d\omega [G_{\mu\nu}^{+}(-\phi, \omega)  G_{\mu\nu}^{-}(\phi, \omega) + G_{\mu\nu}^{+}(\phi, \omega)  G_{\mu\nu}^{-}(-\phi, \omega)],\\
		\mathcal{J}_{DM}=\frac{-J^2}{\pi} {\rm Im}\int_{-\infty}^{E_F} d\omega [i(G_{\mu\nu}^{+}(-\phi, \omega)  G_{\mu\nu}^{-}(\phi, \omega) -G_{\mu\nu}^{+}(\phi, \omega)  G_{\mu\nu}^{-}(-\phi, \omega))].
	\end{align}
\end{widetext}
For QH, $G_{\mu\nu}^{+}(\phi, \omega) = G_{\mu\nu}^{-}(\phi, \omega)$, so from above equations one can easily observe that $\mathcal{J}(DM) = 0$ and we have pure Heisenberg coupling $\mathcal{J}_{H}(E_F, \phi)\textbf{S}_{1}\cdot\textbf{S}_{2}$, with $\mathcal{J}_{H}(E_F, \phi)$ is given by
\begin{align}
	\mathcal{J}_{H}=\frac{-2J^2}{\pi} {\rm Im}\int_{-\infty}^{E_F} d\omega [G_{\mu\nu}(\phi, \omega)  G_{\mu\nu}(-\phi, \omega)].\label{eq:JH}
\end{align}
For simplicity, we consider the spins to be on the same sub-lattice. After some algebra, the integrand becomes
\begin{widetext}
	\begin{align}
		{\rm Im}[G_{AA}(\phi, \omega)  G_{AA}(-\phi, \omega)]=& \sum_{m,n}\left(\frac{\alpha}{2\pi R}\right)^2f_{AA}^{m}(\omega')f_{AA}^{n}(\omega')\Big[(\omega' - m -1/2)(\omega' - n - 1/2) \sin((m-n)\phi)\nonumber\\ +& \eta'^{2}\sin((n-m)\phi) - \eta'(2\omega' - m - n -1)\cos((m-n)\phi)\Big],
	\end{align}
\end{widetext}
with $f_{AA}^{m}(\omega') = C_{AA}^{m}/((\omega' -(m+1/2))^{2} + \eta'^{2})$, $\omega'=\omega\alpha$. Noting that the first two terms in the above sum are odd functions $m$ and $n$,
\begin{align}
	\mathcal{J}_{H}=&\frac{2J^2\alpha}{4\pi^3 R^2}\int_{-\infty}^{E_F \alpha} d\omega' \sum_{m,n} f_{AA}^{m}(\omega')f_{AA}^{n}(\omega')\nonumber\\
	&\times[ \eta'(2\omega' - m - n - 1) \cos((m-n)\phi)].\label{eq:JHfinal}
\end{align}
numerically evaluating the summation we observe that the integrand is independent of $\phi$ and $\mathcal{J}_H$ is positive, dictating antiferromagnetic RKKY coupling between the two impurities placed on the same sub-lattice. The $\phi$ independence is naively because in Eq.~(\ref{eq:JHfinal}) most contribution of the summation comes near $n=m$. This verifies our main result.

\begin{figure}[t]
	\begin{center}
		\includegraphics[width=0.45\textwidth]{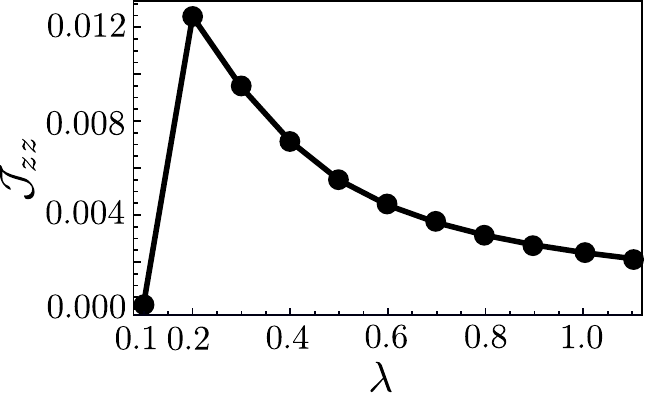}
		\caption{For the hexagonal model in the mein text, the system becomes a QSH state when the spin-orbit coupling $\lambda>\Delta$. For a value of $\Delta=0.1$, the Ising coupling between the spins $\mathcal{J}_{zz}$ decreases with increasing $\lambda$. System size $N_x\times N_y =80\times 16$, $t=1.0$, and impurities are at $(8,Ny)$ and $(30,N_y)$. }\label{fig:lam}
	\end{center}
\end{figure}

For QSH, the inversion symmetry is broken, so $\mathcal{J}_{DM}\neq0$ and also $\mathcal{J}_{I}$, $\mathcal{J}_{H}$ are different. One can observe that $G_{AA}^{+}(\phi, \omega)  G_{AA}^{+}(-\phi, \omega) = G_{AA}^{-}(\phi, \omega)  G_{AA}^{-}(-\phi, \omega)$, so $\mathcal{J}_{I}$ will be same as in equation Eq.~(\ref{eq:JHfinal}). Proceeding as before, the Heisenberg coupling is given by,
\begin{align}
	\mathcal{J}_{H}=&\frac{2J^2\alpha}{4\pi^3 R^2}\int_{-\infty}^{E_F \alpha} d\omega'  \sum_{m,n} f_{AA}^{m}(\omega')g_{AA}^{n}(\omega')\nonumber\\
	&\times[ \eta'(2\omega' - m + n)\cos((m-n)\phi)],\nonumber
\end{align}
here $g_{AA}^{n}(\omega) = C_{AA}^{n}/((\omega' + (n+1/2))^{2} + \eta'^{2})$ and the DM coupling term is given by
\begin{align}
	\mathcal{J}_{DM}=&\frac{2J^2\alpha}{4\pi^3 R^2}\int_{-\infty}^{E_F \alpha} d\omega'   \sum_{m,n}f_{AA}^{m}(\omega')g_{AA}^{n}(\omega')\nonumber\\
	&\times [ \eta'(2\omega' - m + n)\sin((m-n)\phi)].\nonumber
\end{align}
One can check numerically that both integrals depend on $\phi$.

\section{Dependence on $\lambda$}\label{app:V}
In the hexagonal lattice system, larger spin-orbit coupling $\lambda$ implies larger Fermi velocities of the edge states. As the spin-spin coupling is inversely proportional to the Fermi velocity, we expect that the coupling would decrease as the $\lambda$ is increased. Naively, even if the coupling is stronger for smaller $\lambda$, one expects that the coupling would be more susceptible to disorder present in the system. This is confirmed in Fig.~\ref{fig:lam}.


\begin{thebibliography}{99}
\bibitem{analyze}
This can be understood by the fact that for the second order process the connecting states need to traverse the whole perimeter irrespective of the positions of the spins.
\end{thebibliography}
\end{document}